\documentstyle[12pt,aasms4,flushrt,psfig]{article}

\newcommand{\PSbox}[3]{\mbox{\rule{0in}{#3}\includegraphics{#1}\hspace{#2}}}

\newcommand\gammaray{$\gamma$-ray}

\newcommand\psra{PSR~B1046$-$58}
\newcommand\psrb{PSR~B1610$-$50}
\newcommand\psreg{3EG~J1048$-$5840}

\newcommand\asca{{\it ASCA}}

\newcommand\rosat{{\it ROSAT}}

\newcommand\einstein{{\it Einstein}}

\slugcomment{to appear in the Astrophysical Journal}

\begin{document}

\title{\Large\bf {\it ASCA} observations of the young rotation-powered
pulsars \psra\ and PSR B1610$-$50} 

\author{M. J. Pivovaroff\altaffilmark{1} and V. M. Kaspi\altaffilmark{2}}
\affil{Department of Physics and Center for Space Research, 
Massachusetts Institute of Technology, Cambridge, MA 02139, USA}
\authoremail{mjp@space.mit.edu, vicky@space.mit.edu}

\and

\author{E. V. Gotthelf\altaffilmark{3}}
\affil{Columbia Astrophysics Laboratory, Columbia University,
550 W. 120$^{\it th}$ St., New York, NY, 10027, USA}
\authoremail{evg@astro.columbia.edu}

\altaffiltext{1}{mjp@space.mit.edu}
\altaffiltext{2}{Alfred P. Sloan Research Fellow; vicky@space.mit.edu}
\altaffiltext{3}{evg@astro.columbia.edu}

\begin{abstract}

We present X-ray observations of two young energetic radio pulsars,
PSRs~B1046$-$58 and B1610$-$50, and their surroundings, using archival
data from the Advanced Satellite for Cosmology and Astrophysics ({\it
ASCA}).

The energetic pulsar \psra\ is detected in X-rays with a significance
of 4.5$\sigma$.  The unabsorbed flux, estimated assuming a power-law
spectrum and a neutral hydrogen column density N$_H$ of $5 \times
10^{21}$ cm$^{-2}$ is $(2.5 \pm 0.3) \times 10^{-13}$ ergs cm$^{-2}$
s$^{-1}$ in the $2 - 10$ keV band.  Pulsed emission is not detected;
the pulsed fraction is less than 31\% at the 90\% confidence level for
a 50\% duty cycle.  We argue that the emission is best explained as
originating from a pulsar-powered synchrotron nebula.  The X-ray
counterpart of the pulsar is the only hard source within the 95\% error
region of the previously unidentified $\gamma$-ray source
3EG~J1048$-$5840.  This evidence supports the results of Kaspi et al.,
who in a companion paper, suggest that \psra\ is the counterpart to
\psreg.

X-ray emission from \psrb\ is not detected.  Using similar assumptions
as above, the derived $3 \sigma$ upper limit for the unabsorbed $2-10$
keV X-ray flux is $1.5 \times 10^{-13}$ ergs cm$^{-2}$ s$^{-1}$.  We
use the flux limit to estimate the pulsar's velocity to be less than
$\sim$170 km s$^{-1}$, casting doubt on a previously reported
association between PSR~B1610$-$50 and supernova remnant Kes~32.
Kes~32 is detected, as is evident from the correlation between X-ray
and radio emission.  \nocite{kts98a} The \asca\ images of \psrb\ are
dominated by mirror-scattered emission from the X-ray-bright supernova
remnant RCW~103, located 33$^{\prime}$ away.

We find no evidence for extended emission around either pulsar, in
contrast to previous reports of large nebulae surrounding both pulsars.
\nocite{kts98a}

\end{abstract}

\section{Introduction}
X-ray observations of radio pulsars provide a powerful diagnostic of
the energetics and emission mechanisms of rotation-powered neutron
stars.  As magnetic dipole braking slows the pulsar, it loses
rotational kinetic energy at a rate \( \dot{E} = 4\pi^{2}I\dot{P}P^{-3}
\), where $I$ is the moment of inertia of the neutron star, assumed to
be 10$^{45}$ g cm$^{2}$, and $P$ is the rotation period.  Though
pulsars have traditionally been most easily studied at radio
wavelengths, only a small fraction (10$^{-7}$ to 10$^{-5}$) of the
``spin-down luminosity'' $\dot{E}$ manifests itself as radio
pulsations.  Instead, it is believed that a significant fraction of the
luminosity emerges as a relativistic wind of positrons and electrons.

When this wind is confined by the surrounding medium, an observable
synchrotron nebula results.  Measurements of the morphology and
spectrum are essential for determining the
content and energy spectrum of the wind, probing the ambient density,
and understanding the shock acceleration mechanism.  The growing list
of observable X-ray emitting rotation-powered pulsars allows the study
of the properties of the population as a whole.  Young pulsars
constitute a particularly interesting subset to investigate due to
their large spin-down luminosities ($\gtrsim 10^{36}$ ergs s$^{-1}$).

PSR~B1046$-$58 was discovered by Johnston et al. (1992) \nocite{jlm+92}
during a 1400 MHz survey of the Galactic plane for radio pulsars.
\psra\ has $\dot{E} = 2.0 \times 10^{36}$ ergs s$^{-1}$, characteristic
age $\tau_{C} \equiv P/2\dot{P} = 20 \: {\rm kyr}$, and
properties reminiscent of the Vela pulsar.  The dispersion measure (DM)
toward the pulsar is 129 pc cm$^{-3}$.  Using the Taylor \& Cordes
(1993) \nocite{tc93} DM--distance relationship, the distance to
\psra\ is estimated to be $d \approx 3.0$ kpc.  The large $\dot{E}/4
\pi d^{2}$ value (a useful indicator of the detectability of its
putative pulsar wind nebula) make it a strong candidate for X-ray
emission.  Kaspi et al. (1999) have found evidence for $\gamma$-ray
pulsations from an EGRET point source spatially coincident with PSR
B1046$-$58.

PSR~B1610$-$50 was also discovered in the Johnston et al. survey.
\psrb\ is the fourth youngest Galactic pulsar ($\tau_{c} = 7.4$~kyr)
and has $\dot{E} = 1.6 \times 10^{36}$ ergs s$^{-1}$ and ${\rm DM} =
586$~pc~cm$^{-3}$.  \psrb\ occupies an interesting position in the age
phase-space of known pulsars:  its age is between those of the youngest
pulsars ($\tau_{c}$ $\lesssim$ 2~kyr), like the Crab and
PSR~B0540$-$69, and those in their ``adolescence'' ($\tau_{c}$ $\sim$
10-20~kyr), like Vela and PSR~B1046$-$58.  Caraveo (1993)
\nocite{car93} proposed that \psrb\ is associated with the nearby
supernova remnant Kes 32.  However, using the DM-derived distance of
7.3~kpc and assuming the system's age is $\tau_{c}$, the pulsar's
implied transverse velocity is \( v_{t} > 2300 (d/7.3) \) km~s$^{-1}$, 
significantly higher than the mean pulsar
velocity (e.g. \cite{ll94}).  If it is traveling with so high a speed,
one might expect strong ram-pressure confinement of its wind and a
correspondingly high X-ray luminosity from the resultant synchrotron
emission (see e.g.  \cite{che83}; \cite{wlb93}; \cite{fsp96}; \cite{rcy97};
\cite{wg98b}).

Table~1 presents information
for \psra\ and \psrb\, including position, spin parameters, and
derived quantities (i.e. spin-down luminosity and characteristic age).

\placetable{tab1}

Previous X-ray work on PSRs B1046$-$58 and B1610$-$50 has been limited
in scope.  Becker \& Tr\"{u}mper (1997) \nocite{bt97} published a
\rosat-band (0.1$-$2.4 keV) luminosity for \psra\ of $4 \times 10^{32}$
ergs s$^{-1}$, but gave no further analysis.  Kawai, Tamura, \& Saito
(1998) presented \asca\ data on both pulsars, but restrict their
attention to a single Gas Imaging Spectrometer (GIS) image of each
pulsar.  For both PSRs B1046$-$58 and B1610$-$50 they reported the
detection of a large nebula (tens of arcminutes) associated with each
pulsar.  This paper undertakes a detailed analysis of the archival
\asca\ data, with an emphasis on image analysis.

\section{Observations}
\asca\ (\cite{tih94}) observed PSR B1046$-$58 on 1994~January~27 and
PSR B1610$-$50 on 1994~March~25.  We present an analysis of the data
obtained from the public archive. For both observations, data were
taken with all four imaging spectrometers, each in the focal plane of
its own foil mirror:  two Solid State Imaging Spectrometers (SIS-0,
SIS-1) employing charge coupled devices (CCDs), and two Gas Imaging
Spectrometers (GIS-2, GIS-3) employing gas scintillation proportional
counters.  These spectrometers offer moderate energy ($\sim$$5\%$) and
imaging ($\sim$$2^{\prime}$) resolution in their $\sim$$1-10$ keV
energy band-pass.  The SIS has superior imaging and spectral
capabilities, while the GIS has a higher effective area above $\sim$2
keV and a greater net observation time than the SIS.  To facilitate
pulsation searches, GIS data were collected in the highest time
resolution configuration ($0.488$ or $3.906$~$\rm{ms}$ depending on
data acquisition rate). SIS data were acquired in 4 CCD mode with 16~s
integrations (PSR~B1046$-$58) and 2 CCD mode with 8~s
integrations (PSR~B1610$-$50) using a combination of FAINT and BRIGHT
modes (see the \asca\ Data Analysis Guide for details).  The data were
filtered to exclude times of high background contamination using the
standard ``REV2'' screening criteria. This rejects time intervals of
South Atlantic Anomaly passages, Earth block, bright Earth limb in the
field-of-view, and periods of high particle activity.  The resulting
effective observation times per single detector are 18~ks (GIS) and
15~ks (SIS) for PSR~B1046$-$58, and 11~ks (GIS) and 8.4~ks (SIS)
for PSR~B1610$-$50.

\section{PSR~B1046$-$58}
\subsection{Image Analysis}  
\subsubsection{\asca\ Data}
\label{section:ascadata}

Flat-fielded images were generated by aligning and co-adding
exposure-corrected images from the pairs of instruments.  Exposure maps
were generated with ASCAEXPO, \asca\ software which uses the satellite
aspect solution, instrument map (GIS), chip alignment, and hot pixel
map (SIS) to determine the exposure time for each sky image pixel. The
exposure correction was highly effective in removing the GIS
instrumental structure due to the window support grid.  Figure~1a
displays the resultant smoothed broad-band ($0.8-10$~keV) image for the
GIS.  The image reveals emission confined to a slightly oval
$\sim$$4^{\prime} \times 7^{\prime}$ region, elongated along the
direction parallel to declination.  Though the statistics are limited,
the emission appears to be concentrated in regions near the top and
bottom of the oval.

The pulsar location, determined by radio interferometric measurements
made by Stappers et. al (1999), is marked by a cross and lies near the
bottom of the emission region.  The dashed square indicates the region
shown in SIS images (Figures~1b$-$1d).  The two ellipses represent the
95\% and 99\% positional error boxes of the \gammaray\ source
3EG~J1048$-$5840 (Hartman et al. 1999; see below for additional discussion).
Figure~1b displays the central region of the broad-band ($0.4-10$~keV)
image for the SIS, while Figure~1c and Figure~1d show the soft-band
($0.4-2$~keV) and hard-band ($2-10$~keV) images for the same region.  A
cross marks the location of \psra\ and the arcs represent the errors
ellipses for 3EG~J1048$-$5840.  Due to its superior spatial
performance, the SIS resolves the smooth emission seen by the GIS into
four point sources (hereafter called Src~1, Src~2, Src~3, and Src~4)
possibly situated in a region of faint diffuse emission.

The number of detected counts at the pulsar position is too small to
fully resolve the familiar cross pattern of the X-ray telescope (XRT)
PSF, as the morphology is dominated by Poisson fluctuations (see Hwang
\& Gotthelf 1997 \S 2.2 for a discussion of the significance of peaks
in similarly processed \nocite{hg97} images).  To estimate the
significance of the detections, we ignore the complexities of the
\asca\ point-spread function and compare the number of photons
collected from a small aperture centered on the source with that from a
$12^{\prime} - 18^{\prime}$ diameter concentric annulus.  The
relatively small number of counts available in the source region makes
the approximation reasonable.  Unfortunately, the close proximity of
the four sources complicates this analysis.  An optimally
sized\footnote{Here, optimal refers to an aperture that maximizes the
number of source counts captured within the extraction region, relative
to the background contribution.  See e.g. Gotthelf \& Kaspi (1998).
\nocite{gk98}} 4$^{\prime}$ diameter aperture captures the majority of
flux from the pulsar position as well as an undetermined amount of flux
from neighboring sources, resulting in an overestimation of the
pulsar's putative X-ray emission.  Use of a smaller 2$^{\prime}$
diameter aperture eliminates the contamination from the neighbors but
neglects the flux in the broad wings ($\sim$3$^{\prime}$) of the XRT
PSF.  Rather than artificially inflating the significance of a
detection, we employ the 2$^{\prime}$ diameter aperture with the
understanding that our calculations may underestimate the strength of a
source.  Using the formalism outlined by Gotthelf \& Kaspi (1998),
\nocite{gk98} we estimate the detection significance (the
signal-to-noise ratio, $S/N$) with an expression that accounts for both
the source and background variance.

Table~2 presents relevant information for the four sources detected by
both SIS detectors, including source positions, count rates, detection
significance $\sigma$, and hardness ratio $H$, where \( H \equiv \rm
counts(2-10~{\rm keV})/{\rm counts}(0.4-2~{\rm keV}) \).  Src~1 is
relatively hard ($H = 0.70 \pm 0.30$) and is strongly detected with a
significance of 4.5$\sigma$ (44 background subtracted counts).  Src~2
is primarily soft ($H = 0.55 \pm 0.18$) and is strongly detected with a
significance of 6.1$\sigma$ (69 background subtracted counts).  Src~3
is hard ($H = 1.55 \pm 0.82$) and has a significance of 3.9$\sigma$ (38
background subtracted counts), while Src~4 is the hardest source ($H =
2.15 \pm 1.70$) and has a significance of 3.4$\sigma$.  In an attempt to reduce
the influence of the diffuse emission and contamination from
neighboring sources, centroid positions for Src~1, Src~3, and Src~4
were determined from analysis of the hard-band image in Figure~1d and
the position for Src~2 was determined from the soft-band image in
Figure~1c.  Combining the $\sim$15$^{\prime\prime}$ centroid uncertainty for
each source with the $\sim$20$^{\prime\prime}$ revised pointing
uncertainty (\cite{got96}) for \asca\ results in an overall source position 
uncertainty of order $\sim$25$^{\prime\prime}$.

Src~1 lies 20$^{\prime\prime}$ from the radio position
of the pulsar, within the positional errors.  Assuming
approximately 4 sources per square degree with comparable flux to Src~1
(Gendreau, Barcons, \& Fabian 1998) \nocite{gbf98} and a SIS spatial
resolution of 3$^{\prime}$, we estimate the probability of a chance
superposition of Src~1 with the pulsar's position to be of order
0.008.  Src~2 lies 92$^{\prime\prime}$ away from the radio position,
making it extremely unlikely that it is the X-ray counterpart of the
pulsar.  

Our images differ significantly from those produced by Kawai et al.
(1998) from the same \asca\ data. In particular, we do not find any
evidence for a large ($\sim$$10-20^{\prime}$) nebula around the
pulsar.  Our reanalysis of the GIS data and analysis of the SIS data
provide support for emission from the pulsar (Src~1) that is unresolved
by the \asca\ PSF.  Our disparate conclusions result primarily from the
consideration of the SIS data with its superior spatial resolution,
which shows that the oval shaped region in the GIS image actually
represents emission from the four sources resolved by the SIS.  We have
also employed different procedures in the GIS analysis.  Specifically,
we have used an exposure correction that removes the significant
structure produced by the GIS support grid and we have smoothed the
data with a $3 \times 3$ boxcar function ($\sim$45$^{\prime\prime}$ on 
a side) that approximates the core of the 
PSF\footnote{A similar procedure is also performed on the SIS
data.  The data is first rebinned $\times$4, then smoothed with
a $5 \times 5$ boxcar function ($\sim$30$^{\prime\prime}$ on a side).}.
When a much larger smoothing function is used,
Poisson fluctuations, individual sources, and structure arising from
the support grid (if not accounted for) can be blended into an apparent
large, diffuse region of emission.

\placetable{tab2}

\subsubsection{\rosat\ Data}
\label{section:rosatdata}
\psra\ was also observed on 1996~March~8 by the HRI (High Resolution
Imager) onboard \rosat\ for 23~ks.  Several point sources are clearly
detected in the HRI FOV, but none is coincident with the position of
the radio pulsar.  The upper limit for a source at the radio position
is $<$$1 \times 10^{-3}$~cps.  This result is in disagreement with the
detection of \psra\ with the HRI reported by Becker \& Tr\"{u}mper
(1997) \nocite{bt97}.  A source is present, however, at $10^{\rm h}
48^{\rm m} 13.^{\rm s}0$, $-58^{\circ} 30^{\prime} 44^{\prime\prime}$
(J2000), 80$^{\prime\prime}$ away from the pulsar's radio position.
This \rosat\ source is only 19$^{\prime\prime}$ away from SIS Src~2;
given the apparently soft spectrum of Src~2 and the 25$^{\prime\prime}$
positional uncertainty of the \asca\ sources, suggests that they are
the same source.

To try to identify this source (SIS Src~2), we searched several optical
catalogues.  The only coincident source, located
$\sim$$8^{\prime\prime}$ from the \rosat\ position, was found in the
Digital Sky Survey.  Using the photometric calibration provided for the
UK Schmidt Camera, we estimate a V magnitude of $\sim$13.4.  Positive
identification of this optical source with the X-ray source detected by
both \rosat\ and \asca\ requires spectroscopic data not presently
available.

There are no other \rosat\ sources spatially coincident with the
remaining three SIS sources, which is not surprising given their harder
spectra.  Furthermore, no extended or diffuse emission is seen in the
HRI data.  This suggests that the faint diffuse emission seen in
the SIS data may not be physically significant.  One possible origin of
this diffuse structure could be a blending of Poisson fluctuations
with emission from the closely grouped point sources scattered by the
broad PSF of the \asca\ mirrors.

\subsection{Flux Estimation}
While the observations allow detection of the four sources, the low
statistics prevent useful spectral analysis.  However, the hardness
ratio $H$ of the SIS sources and the lack of \rosat\ counterparts for all
but Src~2 give some information about the sources.  Srcs~1,
3, and 4 must either be absorbed non-thermal sources or thermal sources
with temperature of at least several keV.  Src~2 appears to be either
an intrinsically soft thermal source with temperature on order of 50~eV
or is a non-thermal source with a very steep power law that is 
undetectable above 2~keV in the SIS.  

To extract a flux estimate for the X-ray counterpart of the pulsar, we
assume a spectral shape and adjust the overall normalization to match
the count rate\footnote{Here, the count rate is that determined in
$\S$3.1, adjusted to account for the flux in the broad XRT wings that
falls outside the extraction aperture.} of Src~1.  The canonical
synchrotron nebula spectrum is characterized by a power law with photon
index $\alpha = 2$, where $N(E) \propto E^{-\alpha}$ (see e.g.
\cite{sw88} and Becker \& Tr\"{u}mper 1997).  The neutral hydrogen
column density $N_{H}$ can be constrained by combining the Seward \&
Wang (1988) approximation of 10 neutral hydrogen atoms per free
electron with the DM or by using the HI maps of Dickey \& Lockman
(1990). \nocite{dl90}  The former yields $N_{H} = 4.0 \times 10^{21}$
cm$^{-2}$, the latter yields $N_{H} < 1.4 \times 10^{22}$ cm$^{-2}$.
We adopt a value between these two rough estimates of $N_{H} = 5 \times
10^{21}$ cm$^{-2}$.  After folding the spectral model through the
instrument (SIS$+$XRT) response, we obtain an unabsorbed $2-10$ keV
flux of $(2.5 \pm 0.3) \times 10^{-13}$ ergs cm$^{-2}$ s$^{-1}$.
Folding these parameters through the \rosat\ instrument response yields
an expected HRI count rate of $(2.3 \pm 0.3) \times 10^{-3}$~cps, in
rough agreement with the upper limit ($1 \times 10^{-3}$~cps)
calculated above.  The assumed spectral model and $N_{H}$ also agree
well with the observed hardness ratio.

\subsection{Timing}
We carried out a timing analysis for \psra\ using the combined data
from the two GIS detectors.  We selected events in the $2-10$ keV band
from a 4$^{\prime}$ diameter aperture centered on Src~1, using data
acquired at the high and medium data rates only. 
A total of 472 events, a large fraction ($\sim$60\%) of which are
due to the background, were folded using an ephemeris
obtained from radio timing observations of \psra\ at the 64-m Parkes
radio telescope in New South Wales, Australia.  
Table~1 contains the ephemeris.
As the putative pulse shape is unknown, we employed the $H$-test
(\cite{dej94}) to search for pulsations.  For a duty cycle $\delta =
0.5$, the 3$\sigma$ upper limit to the pulsed fraction is 0.31.  For
increasingly sharper pulse shapes of $\delta = 0.3$ and 0.1, the
3$\sigma$ upper limits are 0.22 and 0.12, respectively.  The absence of
pulsations from the GIS data is consistent with the work of Saito
(1998). \nocite{sai98}

\section{PSR~B1610$-$50}
\subsection{Image Analysis}
Flat-fielded images were generated using the same prescription given in
$\S$3.1.  Figure~2a displays the resultant smoothed broad-band
($0.8-12$ keV) image for the GIS.  A cross marks the location of the
pulsar determined from radio interferometric measurements
(\cite{sgj99})\footnote{The interferometric pulsar position differs from the
published catalog value by 57$^{\prime\prime}$ (Taylor et al. 1995). 
\nocite{tmlc95}  No signal is present
at either position.}.  The dashed rectangle shows the SIS FOV.  The
black contours are an overlay of 843 MHz MOST observations of the
supernova remnant Kes~32 (\cite{wg96}).  The flux in the lower left
quadrant results from scattered emission from the X-ray bright
supernova remnant RCW~103, located 33$^{\prime}$ from the GIS optical
axis.  The scattered intensity has the gradient and shape expected when
the 12$^{\prime}$ extent of the SNR is folded through the broad wings
of the {\it ASCA} XRT (\cite{gph97}).  Examination of the soft-band
($0.8-2$ keV) and hard-band ($2-12$ keV) images in Figures~2b and 2c
reveals that the contamination is largely confined to $ E < 2$~keV, due
to the intrinsic spectral nature of RCW~103 and the decrease in the XRT
scattering as a function of increasing energy.  The SIS images have the
same properties as the GIS images; for brevity, we only present the
hard-band ($2-10$ keV) image in Figure~2d.

We again ignore the complexities of the PSF and search for emission
from the pulsar by comparing the number of photons collected from an
(optimal) 4$^{\prime}$ diameter aperture centered on the radio location
and with those collected from a 6$^{\prime} - 11^{\prime}$ diameter
concentric annulus.  By restricting our search to $E > 2$~keV, we
greatly decrease the amount of scattered emission from RCW~103.  Our
choice of background annulus avoids emission from Kes~32 and roughly
contains the same amount of scattered flux as the source aperture
region, allowing a reliable significance calculation.  No emission was
detected by either the GIS or SIS, and the combined detection
significance is below 2$\sigma$.

As first noted by Kawai et al. (1998), the \asca\ observation 
provides the first X-ray detection of the supernova remnant Kes~32.  The
low statistics prevent a detailed comparison of the X-ray emission
with the elongated, shell-like radio morphology.  To first order, the
X-ray flux traces the radio intensity particularly along the
western rim, as is evident in both the SIS and GIS images.  
The absence of X-ray emission from the direction of PSR~B1610$-$50
contradicts a previous report of a large nebula powered by the pulsar
(\cite{kts98a}).  Contamination from the scattered RCW~103 emission
and effects of smoothing with a function larger than the size of
the \asca\ PSF (see the discussion in $\S$\ref{section:ascadata})
can account for the discrepancy.

Two additional sources are also present in the GIS data.  At the top of
the GIS FOV, the \einstein\ source 2E~1611.1$-$5018, a low-mass X-ray
binary with J2000 coordinates $16^{\rm h}14^{\rm m}54^{\rm s}$,
$-50^{\circ}26^{\prime}21^{\prime \prime}$,  is clearly visible.  This
source is detected in both the hard and soft bands and was also
detected by {\it ROSAT}.  The second source, located south of the
pulsar position and just outside the SIS FOV, is only seen above
2~keV.  Data from both GIS detectors provide a 4.1$\sigma$ detection
(47 background-subtracted counts).  Its J2000 coordinates are {\rm
$16^{\rm h}14^{\rm m}18^{\rm s}$}, $-50^{\circ}56^{\prime}43^{\prime
\prime}$, with a position uncertainty of $\sim$$1^{\prime}$.   The
source-like enhancements along the south-eastern edge of the GIS FOV
result from scattered flux from RCW~103 and image processing
artifacts.

\subsection{Flux Estimation}
The non-detection of \psrb\ can be used to place an upper limit on the
flux from the pulsar.  Starting with the observed background
rates\footnote{The background count rate is for an extraction region of
radius $2^{\prime}$.} of $8.2 \times 10^{-3}$ cps, we derive a
3$\sigma$ upper limit on the pulsar's count rate.  We restrict our
analysis to the GIS data, as these have a larger field of view for
background estimation.  We again use the canonical synchrotron nebula
spectral model to estimate the flux and constrain the column density
following the approach taken in $\S$3.2.  We adopt a value of $N_{H} =
2 \times 10^{22}$ cm$^{-2}$, consistent with the Seward \& Wang
estimate and the Dickey \& Lockman upper limit.  After folding the
spectrum through the appropriate instrument response (GIS$+$XRT), we
calculate a 3$\sigma$ upper limit to the unabsorbed $2-10$ keV flux of
$1.5 \times 10^{-13}$ ergs cm$^{-2}$ s$^{-1}$.

\section{\bf Discussion}
The importance of the detection of weak emission from \psra\ and the
non-detection of \psrb\ is most readily understandable in the context
of the growing body of work on the X-ray properties of young ($\tau_{c}
=P/2\dot{P} < 10^{5}$ yr) rotation-powered neutron stars.  More than
twenty of these objects have been detected, with three distinct physical
processes responsible for the observed X-ray flux.  

Thermal emission can result either from the initial cooling of a
young neutron star (e.g. \cite{oge95a} and references therein) 
or from polar-cap reheating in older
pulsars (e.g. \cite{bt93a}).  If the heated cap sweeps across our line
of sight, this emission may also be pulsed, as in the case of the Vela
pulsar, which has an 11\% pulsed fraction in the 0.1$-$2.4 keV band
(\cite{ofz93}), or PSR~B1055$-$52, which has a similar pulsed fraction
that varies with energy in the 0.1$-$10 keV band (\cite{of93};
\cite{gcf+96}).  The blackbody emission is independent
of the spin-down luminosity $\dot{E}$.

Non-thermal magnetospheric emission, produced either in the polar cap or
outer gap regions, is responsible for the classic ``pulsar
phenomenon,'' characterized by sharp pulsations of high pulsed
fraction.  The most famous example is the Crab pulsar, whose pulsed X-ray
spectrum is characterized by a power law with photon index 2 (\cite{ts74}).
The energy for pulsed magnetospheric emission originates from the
spindown and is seen from pulsars having high $\dot{E}$ 
at the extremes of the age distribution, from young ($\tau_{c}
< 10^{4}$ yr) pulsars like the Crab, PSR~B1509$-$58 and the two LMC
pulsars PSR~B0540$-$69 and PSR~J0537$-$6910, to old ($\tau_{c} >
10^{7}$ yr) millisecond pulsars like PSR~J0437$-$4715 and
PSR~B1821$-$24  (\cite{sh82}; \cite{shh84}; \cite{mgz+98};
\cite{bt93a}; \cite{skk+97}).

The pulsar can have an associated synchrotron nebula, the
observable evidence for the channeling of the vast store of rotational 
kinetic energy into a wind of relativistic particles.  Such nebulae arise
when the pulsar wind is confined and subsequently shocked by the
surrounding ISM, e.g. the nebulae produced by the Crab, Vela and
PSR~B1509$-$58 (\cite{bal85} and references therein; 
\cite{hgsk85} and \cite{oz89}; \cite{shss84}, \cite{bb97} and references
therein). 

\subsection{PSR~B1046$-$58}
\label{sec:1046disc}
For an age of 20~kyr, cooling models (see e.g. \"{O}gelman 1995)
predict thermal emission from PSR B1046$-$58 to have an effective
surface temperature of at most $kT \approx  120$ eV and a maximum
bolometric luminosity of $2.3 \times 10^{33}$ ergs s$^{-1}$.  Assuming a
10 km neutron star radius, a 3 kpc distance and $N_{H}
\approx  5 \times 10^{21}$ cm$^{-2}$, the \asca\ count rates should be
no higher than $8\times10^{-3}$ cps (SIS) and $2\times10^{-3}$ cps
(GIS).  While these rates are comparable to the observed rates from the
pulsar direction (refer to Table~2), the predicted count rate
should fall to undetectable levels above 1.5 keV, in contradiction with
the observations.  Thus, cooling thermal emission cannot produce the
observed flux from PSR B1046$-$58.

The apparently hard spectrum of the radiation suggests a non-thermal
origin, either from the magnetosphere or from a synchrotron nebula.
Magnetospheric emission is strongly pulsed, and given the upper limits
on pulsations from the GIS data, it is extremely unlikely that
magnetospheric emission contributes any significant fraction of the
flux.  Deeper observations could reveal pulsations arising from either
the magnetosphere or the modulation of thermal, surface emission.

The most probable source of emission is synchrotron radiation powered
by a relativistic pulsar wind.  The most famous pulsar wind nebula
surrounds the Crab pulsar (\cite{rg74}; \cite{kc84}; \cite{ec87};
\cite{ga94}).  In pulsar wind nebulae, the wind of relativistic
electrons and positrons (and possibly heavy ions, \cite{hagl92}) are
confined, accelerated at the reverse shock, and radiate synchrotron
emission.  Using the flux range calculated in $\S$3.2  the $2-10$ keV
luminosity is $L_{x} = (2.7 \pm 0.3) \times 10^{32} d_{3.0}^{2}$ ergs
s$^{-1}$, where $d_{3.0}$ is the distance in units of 3.0~kpc.  The
conversion efficiency $\epsilon$ of the spin-down luminosity $\dot{E}$
into {\it ASCA}-band emission is $\epsilon = (1.3 \pm 0.1) \times
10^{-4}$.

The $\sim$3$^{\prime}$ broad wings of the \asca\ PSF severely limit
detailed morphological studies of the synchrotron nebula. As the
emission appears consistent with a point source, we can place a
conservative 3$^{\prime}$ diameter upper limit on its angular extent.
The limited spectral resolution also prevents us from considering
whether the faint extended emission surrounding the four SIS sources
could be the supernova remnant associated with PSR B1046$-$58, a
reasonable speculation given the pulsar's apparent youth.  Ultimately,
the nature of the extended emission and its possible relation to
\psra\ will only be resolved through observations with sufficiently
high spatial and spectral resolution that would, for example, allow the
detection of emission line features commonly found in other young
supernova remnants.

\subsection{3EG~J1048$-$5840}
On a list ranked by $\dot{E}/d^2$, a parameter that has proven to be an
excellent indicator of $\gamma$-ray detectability, \psra\ is the ninth,
with six of the eight sources higher being $\gamma$-ray pulsars.
\psra\ thus represents an excellent candidate for observable
high-energy $\gamma$-ray emission.  Indeed, in a companion paper, Kaspi
et al. (1999) \nocite{klm+99} suggest an association between \psra\ and
the unidentified high energy $\gamma$-ray source \psreg\ based on the
detection of significant pulsations from the $\gamma$-ray source at the
radio pulsar period.

Identifying X-ray counterparts to unidentified EGRET sources is a
useful way of significantly reducing the uncertainty in the position of
the $\gamma$-ray source, on the assumption that the source spectrum
extends into the X-ray band, true for both blazars and pulsars (e.g.
\cite{ktm98}; \cite{bt97}).
In Figure 1a, we show the 95\% and 99\% confidence
spatial contours of 3EG J1048$-$5840 overlayed on the broad-band GIS
image (R. Hartman, personal communication).
Fortunately, the archival \asca\ observation covers the entire $\gamma$-ray
error box.  The three detected \asca\ sources discussed above ($\S$3.1) are
the only significant X-ray sources in the field; one or more of them is
therefore probably the source of the $\gamma$-ray emission.

The {\it ASCA} source nearest (within the 95\% contour) the best-fit
3EG J1048$-$5840 position is Src~1, which we have identified with
\psra.  It therefore represents the most likely counterpart to 3EG
J1048$-$5840, and supports the evidence presented by Kaspi et al.
(1999).  Src~2, outside the 95\% contour but within the 99\% region,
was shown ($\S$3.2) to have a soft spectrum, hence is unlikely to be
the $\gamma$-ray source counterpart.  The unidentified hard-spectrum
Src~3 lies well outside the 95\% contour but just within the 99\%
contour, so we cannot formally preclude its being the counterpart.
However, were it the counterpart, it would most likely have to be a
{\it second} young, energetic pulsar in the field, as it has no obvious
radio counterpart, hence cannot be a blazar, since all known
$\gamma$-ray emitting blazars are bright radio sources
(\cite{msm+97}).

\subsection{PSR~B1610$-$50}
The non-detection of \psrb\ obviously prevents a study of the emission
characteristics of the pulsar.  However, derived upper limits are
important for studying and understanding the X-ray properties of young
rotation powered pulsars as a population, particularly given this
pulsar's place in age-space between the youngest Crab-like pulsars and
the older Vela-like pulsars.

At an age $\tau = 7.4$~kyr, the \"{O}gelman (1995) cooling model
predicts thermal emission from the surface of the neutron star with a
characteristic temperature $kT \approx 130$ eV.  With the large column
density towards the pulsar and its distance,  though, the absorbed flux
in the $2-10$ keV band would be $F_{x} = 2.1 \times 10^{-17}$ ergs
s$^{-1}$ cm$^{-2}$, orders of magnitude below the \asca\ detectability
threshold.  Pulsed magnetospheric emission is also expected to be
present, given that pulsars with similar $\dot{E}$ exhibit pulsed
non-thermal emission.  However, as the total flux in the pulsed
radiation is usually less than that from a synchrotron nebula, we only
used the non-thermal X-ray emission expected from the nebula in our
calculation of an upper flux limit in $\S$3.3.  For a flux limit $F_{x}
< 1.5 \times 10^{-13}$ ergs s$^{-1}$ cm$^{-2}$ and a distance of 7.3
kpc, the $2-10$ keV luminosity is $L_{x} < 9.6 \times 10^{32}
d_{7.3}^{2}$ ergs s$^{-1}$.  The conversion efficiency $\epsilon$ of
the spin-down luminosity $\dot{E}$ into {\it ASCA}-band emission is
$\epsilon < 6.1 \times 10^{-4}$.

If we assume that \psrb\ has an X-ray emitting synchrotron nebula, it
is likely to be confined by ram pressure, as there is no evidence from
X-ray or radio observations of a confining shell (i.e. SNR) around the
pulsar (\cite{gre98}).  Relying on the work of Arons \& Tavani (1993)
\nocite{at93} and others, Gotthelf \& Kaspi (1998 and references
therein) show that the cooling efficiency for relativistic pairs of
positrons and electrons is:
\begin{equation}
\epsilon \equiv \frac{t_{f}}{t_{s}}  = 3.6 \times 10^{-4} (\frac{\sigma}{0.005})
(\frac{\rho}{1 \: \rm H \: \rm atom})^{1/2} (\frac{v}{100 \: \rm km \: \rm s^{-1}})( \frac{\gamma}{10^8}), 
\end{equation}
\noindent

where $t_{f}$ and $t_{s}$ are the time scales for the pulsar wind flow
and synchrotron cooling, $\sigma$ is the ratio of magnetic energy flux
to the kinetic energy flux of the wind ($\sigma \approx 0.005$ for the
Crab pulsar, \cite{kc84}), $\rho$ is the ambient density, $v$ is the
pulsar's velocity, and $\gamma$ is the postshock pair Lorentz factor.
(\cite{tbd+95}).  The bolometric luminosity in synchrotron emission for
the highest energy pairs is \( L_{s} \approx  \epsilon \dot{E} \).  If
\psrb\ has wind properties similar to the Crab pulsar, the velocity
must be $\lesssim 170$ km s$^{-1}$ given the absence of X-ray
emission.  This velocity is inconsistent with an association between
the pulsar and Kes~32.  Unless the pulsar resides in an extremely
underdense region, i.e.  $\rho \lesssim 5.4 \times 10^{-3}$ cm$^{-3}$,
or has very  different wind properties compared to the Crab pulsar,
\psrb\ does not have the transverse velocity suggested by Caraveo
(1993).

\subsection{  {\boldmath $L_{x}-\dot{E}$} relationships}
The luminosity measured for \psra\ and the upper limit derived for
\psrb\ can be considered within the context of two previously
determined empirical correlations between spin down luminosity
$\dot{E}$ and X-ray luminosity $L_{x}$:  the Seward \& Wang (1988)
relationship (hereafter SW88) derived 
from \einstein\ data ($0.2 - 4.0$ keV band),  
\( \log L_{x} = 1.39 \log \dot{E} - 16.6 \), and the Becker \& Tr\"{u}mper
(1997) relationship (hereafter BT97) 
derived from \rosat\ data ($0.1 - 2.4$ keV band),
\( L_{x} = 10^{-3} \dot{E} \).  To estimate the scatter associated with
these models, we have computed the root mean square (RMS), defined here
to be the square root of the variance, using the data presented by the
authors in their papers.  For the \einstein\ model, the RMS scatter is
a factor of $\sim$7, while for the \rosat\ model, the RMS scatter is a
factor of $\sim$4.  For \psra, SW88 predicts
$7.2^{\: 51}_{\: 1.0} \times 10^{33}$~ergs~s$^{-1}$, compared to
the observed 
$(5.0 \pm 0.5) \times 10^{32}$~ergs~s$^{-1}$, while
BT97 predicts $2.0^{\: 8.0}_{\: 0.5} \times 10^{33}$,
compared to the observed $(5.3 \pm 0.5) \times 10^{32}$~ergs~s$^{-1}$.
In both cases, the 
relationships overestimate $L_{x}$ by factors of at least several.  
For \psrb, SW88 predicts
$5.3^{\: 37}_{\: 0.8} \times 10^{33}$~ergs~s$^{-1}$, compared with
the derived upper-limit $1.3 \times 10^{33}$~ergs~s$^{-1}$, while
BT97 predicts $1.6^{\: 6.4}_{\: 0.4} \times 10^{33}$,
compared with the derived upper-limit $1.4 \times 10^{33}$~ergs~s$^{-1}$.
For this pulsar, the derived upper-limits lie within
the range of predicted values.


One possible source of error is the distance derived from the Taylor \&
Cordes (1993) DM-$d$ model.  Distances are uncertain to
$\sim$25$-$50\%, translating to a potential error as a large as a
factor of $\sim$3 when converting flux to luminosity.  Increasing the
distance by $\sim$25$-$50\% leads to a corresponding increase in
$L_{x}$ and helps improve the agreement between model and observation.
The discrepancies between model and observation also suggests that the
current $L_{x}$-$\dot{E}$ relationships, while certainly illustrating a
correlation between spin-down and X-ray luminosity, may
overlook important factors like
the pulsar's velocity or the ISM conditions in its vicinity.

\section{Conclusions}
We have analyzed archival \asca\ X-ray data for pulsars \psra\ and
PSR~B1610$-$50.  We have detected emission from \psra\ with a
4.5$\sigma$\ significance, the emission most likely arising from a
synchrotron nebula.  We estimate the $2-10$ keV unabsorbed flux to be
$F_{x} = (2.5 \pm 0.3) \times 10^{-13}$ ergs cm$^{-2}$ s$^{-1}$,
assuming a distance $d = 3.0$~kpc.  The data show no evidence for
pulsation, with pulsed fraction less than 31\% at the 90\% confidence
limit for a duty cycle of 50\%.  The synchrotron nebula is the only
hard source within the 95\% error ellipse of the previously unidentified
$\gamma$-ray source \psreg\, and one of only two hard sources within
the 99\% error ellipse.  
The X-ray data, coupled with $\gamma$-ray pulsations discovered by 
Kaspi et al. (1999), strengthens the case for identifying \psra\
as one of the few radio pulsars known to emit pulsed $\gamma$-rays.
We detect no emission from \psrb\, and place an upper limit on the
$2-10$ keV unabsorbed flux $F_{x} < 1.5  \times 10^{-13}$ ergs
cm$^{-2}$ s$^{-1}$.  This flux limit argues that the pulsar's velocity
$v \lesssim 170$ km s$^{-1}$, providing evidence against the
association between by \psrb\ and Kes~32 claimed by Caraveo (1993).

The flux-derived luminosities for both pulsars are below those
predicted by the Seward \& Wang (1988) and Becker \& Tr\"{u}mper (1997)
$L_{x}-\dot{E}$ relationship, particularly for \psra.  The differences
may represent the intrinsic limitations of the current empirically
derived $L_{x}-\dot{E}$ relationships, which do not consider parameters
such as ambient density and pulsar velocity.

The unresolved nebula powered by \psra\ and the non-detection from
\psrb\ contradict previous reports by Kawai et al. (1998) of
large ($\sim$$10-20^{\prime}$) nebula associated with these pulsars.
These observations also demonstrate the difficulties associated
with analyzing and interpreting current hard X-ray ($E > 2$ keV)
imaging data.  The upcoming launch of the {\it Chandra X-ray
Observatory}, with its
unparalleled resolution ($\sim$$0^{\prime\prime}.5$) 
and hard-band sensitivity, and {\it XMM} with its
good spatial resolution ($\sim$$8^{\prime\prime}$) and large collecting
area, should improve this situation considerably.

\bigskip

\acknowledgments 
We thank B. Gaensler for providing the radio images of the
\psra\ and PSR~B1610$-$50-Kes~32 fields and for useful discussion.  We
thank M.  Bailes, R. Manchester, and R. Pace for acquiring radio timing
data for \psra\ at the 64-m Parkes radio telescope in New South Wales,
Australia.  We thank R. Hartman for providing information on 3EG
J1048$-$5840 prior to publication.  We thank W. Becker for discussion 
on the analysis of the   \rosat\  observations of \psra.
Finally, we thank
the anonymous referee for useful comments and suggestions that improved
this paper.
We have used the NASA maintained
HEASARC web site extensively for archival data retrieval and subsequent
analysis.  M. Pivovaroff is supported in part by NASA under Contract
NASA-37716.  

\clearpage

\begin{deluxetable}{l r r}
\small
\tablewidth{0pt}
\tablecaption{Astrometric and Spin Parameters for PSRs B1046$-$58 and 
B1610$-$50.  \label{tab1} }
\tablehead{
 \colhead{Parameter} & \colhead{\psra}  & \colhead{\psrb} 
}
\startdata
    Right ascension (J2000)\tablenotemark{a}	& $10^{\rm h} 48^{\rm m} 12.^{\rm s}6$ & $16^{\rm h} 14^{\rm m} 11.^{\rm s}6$ \nl 
    Declination (J2000)\tablenotemark{a}	& $-58^{\circ} 32^{\prime} 03.^{\prime\prime}8$ & $-50^{\circ} 48^{\prime} 01.^{\prime\prime}9$ \nl 
    Period, $P$ (s)\tablenotemark{b,c}			& 0.124  &  0.232  \nl
    Period derivative, $\dot{P}$ (s s$^{-1}$)\tablenotemark{b} &  $95.9 \times 10^{-15}$ & $493 \times 10^{-15}$   \nl
    Epoch of period (MJD)             & 49403.0    &  48658.0 \nl
    Dispersion measure, DM (pc cm$^{-3}$)\tablenotemark{b} & 129.09(1)   & 586(5)  \nl
    Characteristic age, $\tau_{c}$ (kyr)	      & 20.4&  7.4\nl
    Spin-down luminosity, $\dot{E}$ (ergs s$^{-1}$)  & $2.0 \times 10^{36}$  & $1.6 \times 10^{36}$ \nl
    Distance, $d$ (kpc)\tablenotemark{d}   &   3.0         & 7.3   \nl
    Column density, $N_{H}$ (cm$^{-2}$)\tablenotemark{e} & $(0.40-1.4) \times 10^{22}$   & $(1.8-2.2)  \times 10^{22}$  \nl
\enddata
\tablenotetext{a}{Radio positions, from \cite{sgj99}, are uncertain by $< 0.1^{\prime\prime}$.}
\tablenotetext{b}{Data obtained from the Taylor et al. (1995) \nocite{tmlc95} pulsar catalog.  This information is not used for the timing analysis of
PSR~B1046$-$58.}
\tablenotetext{c}{Ephemeris for \psra\ was obtained from radio timing
observations at the 64-m Parkes radio telescope in New South Wales,
Australia.}
\tablenotetext{d}{Derived from the Taylor $\&$ Cordes (1993) DM-distance model.}
\tablenotetext{e}{Lower limits were derived from the Seward \& Wang (1988) estimate of 10 neutral hydrogen atoms per free electron; upper limits were derived from Dickey \& Lockman (1990).}
\end{deluxetable}

\clearpage
\begin{deluxetable}{c c c c c c c c}
\small
\tablewidth{0pt}
\tablecaption{\asca\ SIS detection of \psra~(0.4-10 keV Band).  \label{tab2} }
\tablehead{
\colhead{Source}  & \colhead{Right ascension} & \colhead{Declination} & \colhead{Count Rate} & {Background Rate}  & \colhead{Hardness} & \colhead{Significance} \\
   &\colhead{(J2000)}   & \colhead{(J2000)}  &  \colhead{($ \times 10^{-3}$ cps)}  & \colhead{($ \times 10^{-3}$ cps)}  & \colhead{Ratio, H} & \colhead{($\sigma$)} 
}
\startdata
  Src~1 & $10^{\rm h} 48^{\rm m} 11.^{\rm s}$6 & $-58^{\circ} 31^{\prime} 46^{\prime\prime}$ & \phn3.99$\pm$0.42 & $2.10 \pm 0.04$ & $0.70 \pm 0.30$ & 4.5 \nl
  Src~2 & $10^{\rm h} 48^{\rm m} 15.^{\rm s}$1 & $-58^{\circ} 30^{\prime} 34^{\prime\prime}$ & \phn4.87$\pm$0.43 & $2.18 \pm 0.04$ & $0.55 \pm 0.18$ &  6.1 \nl
  Src~3 & $10^{\rm h} 48^{\rm m} 04.^{\rm s}$4 & $-58^{\circ} 28^{\prime} 26^{\prime\prime}$ & \phn3.62$\pm$0.38 & $2.15 \pm 0.05$ & $1.48 \pm 0.82$ & 3.9 \nl
  Src~4 & $10^{\rm h} 48^{\rm m} 15.^{\rm s}$1 & $-58^{\circ} 28^{\prime} 57^{\prime\prime}$ & \phn3.27$\pm$0.38 & $2.21 \pm 0.04$ & $2.15 \pm 1.70$& 3.4\tablenotemark{a} \nl

\enddata
\tablecomments{The positions for Src~1, Src~3, and Src~4 were derived from the
hard band ($2-10$ keV) image; the position for Src~2 was derived from the
soft band ($0.4-2$ keV) image.  Total uncertainties in the source positions are
$\sim$25$^{\prime\prime}$.
Count rate is the total source
plus background count rate in an aperture centered on the source
position.  Background rate is the count rate in  a
$12^{\prime}-18^{\prime}$ diameter annulus concentric with the source
position, normalized to the source aperture.  Refer to the text 
($\S$3 \& $\S$4) for the
definition of significance and further discussion on all the measured
quantities. The hardness ratio H is defined as counts in the hard band ($2-10$ keV) 
divided by counts in the soft band ($0.4-2$ keV). } 
\tablenotetext{a}{This significance is for the $2-10$ keV band.  The broad-band significance is 3.2$\sigma$.}
\end{deluxetable}

\clearpage

\setcounter{figure}{0}
\refstepcounter{figure}
\PSbox{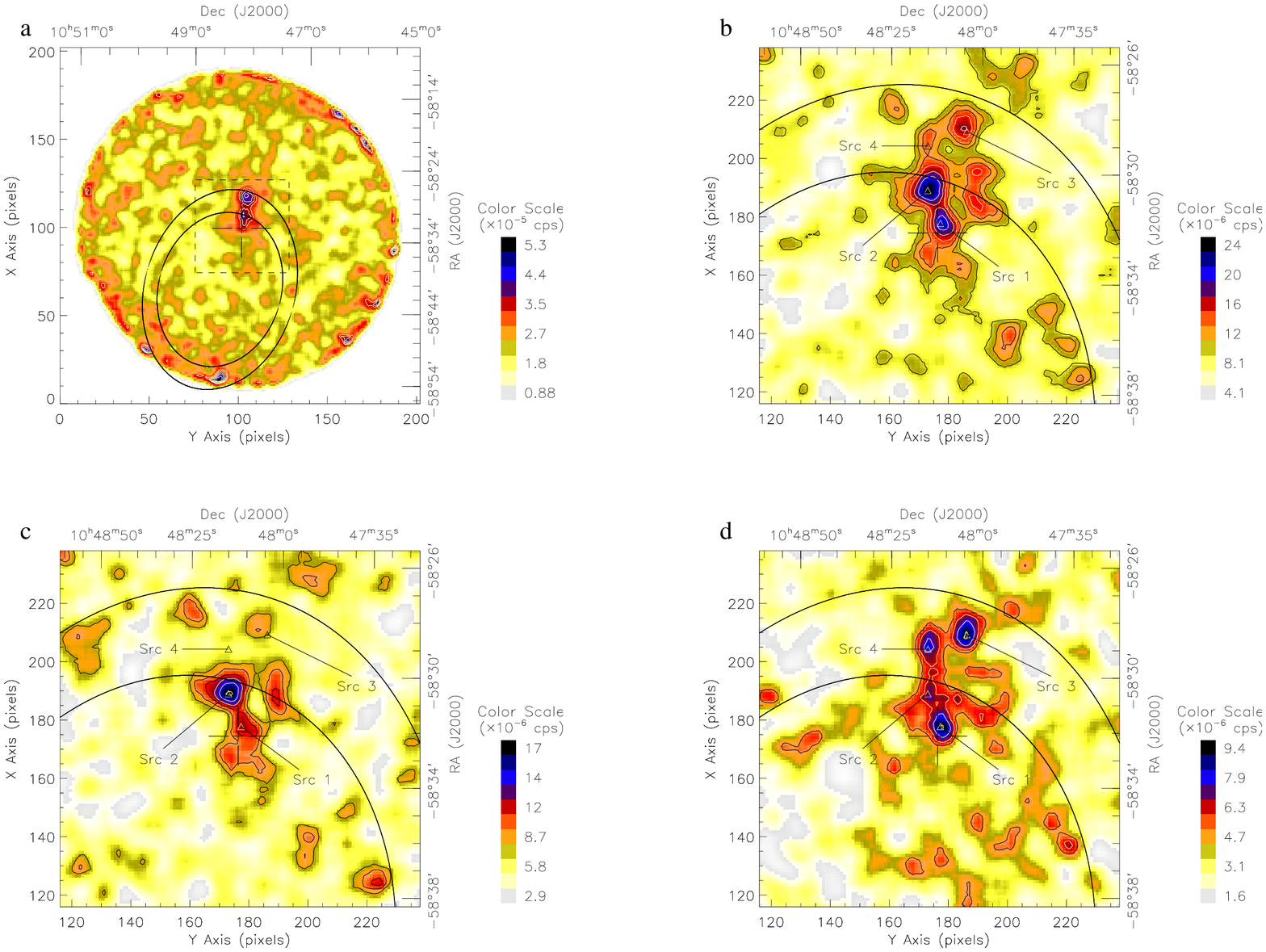 hoffset=450 voffset=-95 hscale=90
vscale=90 angle=90}{6in}{7.8in}{\\\\\small Fig. 1 -- 
{\asca\ images of the \psra\ field: flat-fielded images of the
region around the pulsar, whose location is marked by the cross. a) The
broad band ($0.8-12$ keV) GIS image shows an oval shaped region of
X-ray emission with the pulsar located at its southern tip.  The two
ellipses represent the 95\% and 99\% error boxes for the
\gammaray\ source 3EG~J1048-5840.  The dashed square delineates the SIS
region displayed in b)$-$d).  b) The broad band ($0.4-10$ keV) SIS
image clearly showing the three labeled sources embedded in a diffuse
emission region.  c) The soft band SIS image ($0.4-2$ keV) revealing
the soft, probably thermal nature of Src~2.  Note that Srcs~1, 3, and 4
are very weak in this band.  d) The hard band ($2-10$ keV) SIS image
showing the hard nature of Srcs~1, 3 and 4.  We
identify Src~1, offset 20$^{\prime\prime}$ from the radio position of
PSR~B1046$-$58 and the only source inside the 95\% error circle of the
pulsed \gammaray\ source \psreg, as the synchrotron nebula of
PSR~B1046$-$58.  Contours approximately correspond to the 4$\sigma$,
5$\sigma$, 6$\sigma$, 7$\sigma$, 8$\sigma$, and 9$\sigma$ levels.
Count rates are in units of $10^{-5}$ cps/pixel for the GIS
and $10^{-6}$ cps/pixel for the SIS.
}

\refstepcounter{figure}
\PSbox{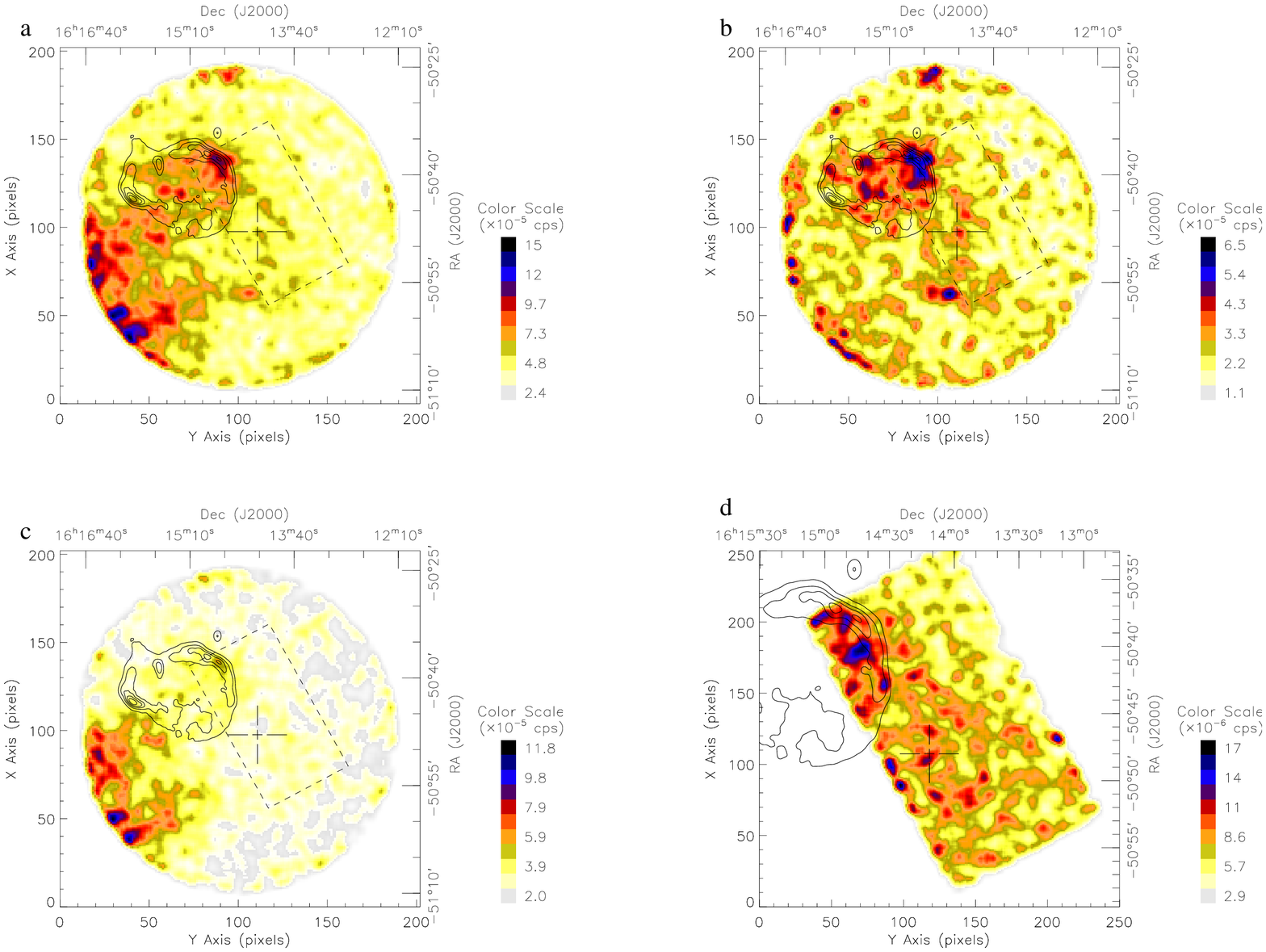 hoffset=450 voffset=-95 hscale=90
vscale=90 angle=90}{6in}{7.8in}{\\\\\small Fig. 2 -- 
{\asca\ images of the \psrb\ field:  flat-fielded images of the
region around the pulsar, whose location is marked by the cross.  The
dashed rectangle represents the SIS FOV, and the contours,
corresponding to 0.04, 0.18, 0.31, 0.44, 0.57, and 0.70 Jy beam$^{-1}$,
are from 843 MHz MOST observations of the supernova remnant Kes~32.  a)
The broad band ($0.8-12$ keV) GIS image of the \psra\ field.  Scattered
emission from the nearby supernova remnant RCW~103 is responsible for
the large flux gradient that begins in the southeast FOV and extends to
th edge of the SIS FOV.  b) The hard band ($2-12$ keV) GIS image shows
enhanced emission that traces the radio emission from Kes~32.  No
significant flux is seen from the pulsar location.  The previously
known \einstein\ source 2E~1611.1$-$5018 is visible at the top of the
FOV, while an unidentified  source is located approximately due south
of the pulsar position.  c) The soft band ($0.8-2$ keV) GIS image
explicitly shows the extent of the scattered emission from RCW~103.
Note the distinct {\it lack} of emission from Kes~32.  d) The hard band
($2-10$ keV) SIS image similarly shows the  correspondence between the
radio contours and X-ray emission and no emission from PSR~B1610$-$50.
Count rates are in units of $10^{-5}$ cps/pixel for the GIS
and $10^{-6}$ cps/pixel for the SIS.
}

\end{document}